\documentclass[12pt,preprint]{aastex}

\slugcomment{to appear in Ap.J.}

\shorttitle{Thermohaline instabilities inside stars}
\shortauthors{Vauclair & Th\'eado}

\begin{document}

\title{Thermohaline instabilities inside stars: a synthetic study including external turbulence and radiative levitation}

\author{Sylvie Vauclair \altaffilmark{1,2,3} and Sylvie Th\'eado \altaffilmark{1,2}}
\affil{Universit\'e de Toulouse; UPS-OMP; IRAP; Toulouse, France
   \and
     CNRS; IRAP; 14 avenue Edouard Belin, F-31400 Toulouse, France
   \and 
     Institut universitaire de France, 103 boulevard Saint Michel, 75005, Paris, France}
\email{sylvie.vauclair@irap.omp.eu}

\begin{abstract}

We have derived a new expression for the thermohaline mixing coefficient in stars, including the effects of radiative levitation and external turbulence, by solving Boussinesq equations in a quasi-incompressible fluid with a linear approximation. It is well known that radiative levitation of individual elements can lead to their accumulation in specific stellar layers. In some cases, it can induce important effects on the stellar structure. Here we confirm that this accumulation is moderated by thermohaline convection due to the resulting inverse $\mu$-gradient. The new coefficient that we have derived shows that the effect of radiative accelerations on the thermohaline instability itself is small. This effect must however be checked in all computations. We also confirm that the presence of large horizontal turbulence can reduce or even suppress the thermohaline convection. These results are important as they concern all the cases of heavy element accumulation in stars. The computations of radiative diffusion have to be revisited including thermohaline convection and its consequences. It may be one of the basic reasons for the fact that the observed abundances are always smaller than those predicted by pure atomic diffusion. In any case, these processes have to compete with rotation-induced mixing, but this competition is more complex than previously thought due to their mutual interaction.

\end{abstract}




\section{Introduction}

During many decades, the theory of thermohaline convection was developed in the framework of oceanography, as it represents one of the major physical factors leading to the large streams of the earth ocean (\cite{Stern60}, \cite{Stern67}, \cite{Turner73}, \cite{Gargett03}). It was not much applied to astrophysics, except for very few papers, like \citet{Ulrich72} and \citet{Kippenhahn80} (hereafter KRT80). At the present time, thermohaline convection is recognized as a major mixing process which occurs in stars in the presence of an unstable $\mu$-gradient associated with a stable temperature gradient, provided that the $\mu$-gradient is not large enough to drive dynamical convection. If a blob of stellar material begins to move down, it exchanges heat with its surroundings more rapidly than particles, and goes on falling, thereby triggering the instability (\cite{Vauclair04} and references therein). 

Various situations lead to thermohaline convection in stellar interiors. They may be classified as follows:

\begin{itemize}

\item{Accretion of heavy matter onto the star, which may come from planetary material \citep{Vauclair04, Garaud11, Theado12} or from an evolved companion in a binary system \citep{Stothers69,Stancliffe07,Thompson08}.}  

\item{Local reduction of the mean molecular weight due to nuclear reactions, for example in Red Giants \citep{Eggleton06,Eggleton08,Charbonnel07,Charbonnel10,Stancliffe10, Denissenkov10}.}

\item{Heavy elements accumulation induced by atomic diffusion \citep{Theado09} or helium accumulation induced by diffusion in a mass loss flux like in main-sequence helium rich stars \citep{Vauclair75}}.

\end{itemize}

The variations with depth of the radiative accelerations on specific elements can lead to their accumulation or depletion in various layers inside the stars, which may have strong consequences on the stellar structure and evolution \citep{Richer00, Richard01}. The accumulations of iron and nickel, which represent important contributors to the opacity in some stellar layers, may lead to extra convective zones, and can even, in some cases, trigger stellar pulsations through the iron-induced  $\kappa$-mechanism \citep{Charpinet96, Pamyatnykh04, Bourge06}. However, in all these previous studies, the thermohaline convection induced by the heavy elements accumulation was not taken into account. First computations of this effect were given in \citet{Theado09}, who found that the accumulation of heavy elements was attenuated by the thermohaline mixing process, but not completely suppressed. The helium settling which occurs simultaneously with the heavy element accumulation induces a stabilizing contribution to the global $\mu$-gradient, which helps keeping some accumulation.

Recent numerical simulations of thermohaline convection (\cite{Denissenkov10}, hereafter D10, and \cite{Traxler11}, hereafter TGS11) brought a new light on these previous computations. As discussed by \cite{Theado12}, the thermohaline mixing coefficients previously used in the litterature have to be revised.
Also, in the computations by \citet{Theado09}, the effect of the radiative acceleration on the falling blobs was not introduced. Finally the influence on the thermohaline convection of an external turbulence, already addressed by \cite{Denissenkov08}, is derived and introduced in the present context. 

In the present paper, we analyse this situation and we give a new derivation of the thermohaline mixing coefficient, including radiative accelerations and/or external turbulence, consistent with the numerical simulations. This new coefficient will be used in the future to compute the evolution of the element abundances inside stars, including the simultaneous effects of atomic diffusion and thermohaline mixing. The introduction of thermohaline mixing in the computations may have important consequences on the stellar structure and evolution.

\section{Modelling thermohaline convection inside stars}

Thermohaline convection is not a real diffusion process. It is a special kind of convection which presents, at least in the ocean case, elongated cells called ``salt fingers". This convection, induced by an unstable chemical stratification, leads to chemical mixing. For this reason, and for simplicity, it is generally modelled in the same way as diffusion, with a mixing coefficient whose expression has to be deduced from physical considerations. In the past, the mixing coefficients proposed for stellar conditions were very uncertain, with unknown normalisation values. For example, the expressions given by \citet{Ulrich72} and KRT80 differ by nearly two orders of magnitude. Recently, numerical simulations like those of TGS11 have led to real improvements, as the mixing coefficient must account for the chemical fluxes obtained from the simulations.  

An important physical parameter in the study of thermohaline convection is the so-called ``density ratio" in oceanography, which may be transposed to the astrophysical case as: 
\begin{equation}
 R_0 = \frac{\nabla_{ad}-\nabla}{|\nabla_{\mu}|}
\end{equation}
where $\nabla_{ad}$ and $\nabla$ are the usual adiabatic and local (radiative) gradients $d \ln T/ d \ln P$, and  $\nabla_{\mu}=d \ln \mu/d \ln P$.

As discussed in \cite{Vauclair04}, in the absence of external turbulence or radiative levitation, thermohaline convection may develop only if:
\begin{equation}
1 < R_0 < \frac{1}{\tau}
\end{equation}
where $\tau$ is the inverse Lewis number, equal to the ratio of the particle diffusivity to the thermal diffusivity.

These two boundaries, which are well known in oceanography, have important physical meanings. For thermohaline convection to occur, the medium must be stable according to the Ledoux criterion (no dynamical convection), but the relative variations of $T$ and $\mu$ inside a falling blob must be large enough, compared to the surroundings, to trigger the instability. This last condition imposes that the ratio of the thermal diffusivity to the particle diffusivity, which induces the variations of these two quantities inside the blob, be larger than the external ratio of the temperature gradient to the particle gradient.

If $R_0$ is smaller than one, the $\mu$-gradient triggers dynamical convection. If $R_0$ is larger than $1/\tau$, the initial $\mu$-gradient is not large enough to overcome the stabilizing effect of the temperature gradient. This means that the expression of the thermohaline coefficient must behave in the right way at these two limits: it must become very large for $R_0$ close to one, and it must vanish for $R_0$ close to $1/\tau$. This behavior is not reproduced by any of the previously used thermohaline mixing coefficients, except for that proposed by D10, as discussed below.

\citet{Ulrich72} and KRT80 proposed to model the thermohaline mixing process in terms of a diffusion coefficient proportional to the inverse $\mu$-gradient, which can simply be written (see \cite{Theado12}):
\begin{equation}
D_{th}=C_t k_T R_0^{-1}
\end{equation}
where $k_T$ is the thermal diffusivity, and $C_t$ an unknown factor including the aspect ratio (length/width) of the fingers. $C_t$ is evaluated as 12 by KRT80, and 658 by \cite{Ulrich72}. In their computations of red giant stars, \cite{Charbonnel07} considered this parameter as free and adjusted it as $C_t = 1000$. Note that this expression does not satisfy any of the two physical limits: $D_{th}$ goes to infinity for $R_0$ = 0 and not $R_0$ = 1 as needed, and it does not vanish for $R_0 = 1/\tau$.

The recent numerical simulations by TGS11 considerably improved our knowledge of thermohaline convection. They found universal scaling laws for the turbulent heat and compositional fluxes, valid for different sets of Prandt and Lewis numbers. Fitting their numerical results by an empirical law, TGS11 gave an expression for $D_{th}$ which vanishes for $R_0 = 1/\tau$, but reaches a finite limit for $R_0 = 1$, which is not correct. Interestingly enough, a close look at their simulation points (their Figure 2) shows that the heat and compositional fluxes do increase for small $R_0$, far above their empirical relation, suggesting that this relation should be revised. 

\cite{Denissenkov10} (D10) derived a thermohaline mixing coefficient using a straightforward linear treatment of Boussinesq equations in a quasi-compressible fluid. He found an expression which nicely satisfies the boundary conditions. As \cite{Ulrich72} and KRT80, he introduced an adjustable parameter, $C_t$, depending on the shape of the convective cells (aspect ratio of the fingers) and tried to constrain it by 2D simulations. Here we prefer using the more recent 3D simulations of TGS11 to adjust this parameter. We find that the D10 expression nicely fits the simulations, including the increase of $D_{th}$ when $R_0$ becomes close to one, which is visible on the simulation but was not introduced in the TGS11 expression:
\begin{equation}
D_{th}=C_t k_T (R_0 - 1)^{-1}(1-R_0\tau)
\end{equation}

Figure 1 presents the behavior of all these coefficients as a function of $R_0$, in stellar conditions. As an example, they have been computed for a $ 1.7 M_{\sun}$, 403 Myr model. Here the $1/\tau$ value, which is related to the rapid decrease of $D_{th}$ for large values of $R_0$, is $6.8. 10^8$, which is larger than the values generally given in the literature (e.g. TGS11 or \cite{Theado12}). The basic reason for this difference is that this parameter is generally computed with helium as the diffusing element, whereas here we take into account the fact that the diffusive element is iron, which diffuses much more slowly than helium (C.f. Figure 2). From all the considerations discussed above, the coefficient proposed by D10 is clearly the best one, as it satisfies the results of the 3D simulations, as well as the boundary conditions. However, this coefficient does not take into account external turbulence, neither the effect of the radiative acceleration on the falling blobs, which may be important when the instability is induced by heavy element accumulation due to radiative levitation. In the following, we derive a new thermohaline coefficient in the same way as D10, including these two effects.

\begin{figure*}
\center
\includegraphics[width=0.5\textwidth]{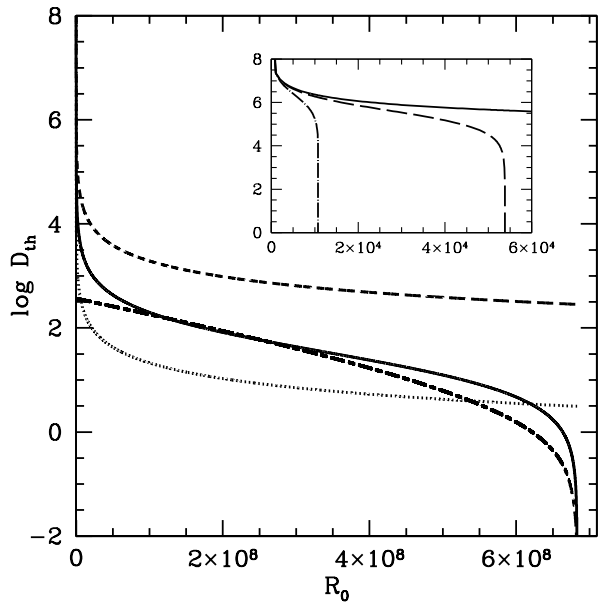}
\caption{Examples of various prescriptions for the thermohaline diffusion coefficients as computed in a $1.7 M_{\sun}$, 403 Myr model, with a $1/\tau$ value of $6.8. 10^8$. In the main frame, the dotted curve represents the KRT80 coefficient, with $C_t$ = 12, as a function of $R_0$, the dashed curve the \cite{Charbonnel07} coefficient, with $C_t$ = 1000, the dotted-dashed curve represents the TGS11 expression and the solid curve the D10 coefficient, adjuste here with $C_t$ = 120 to fit the TGS11 simulations. In the small frame, the solid curve is a zoom of the D10 coefficient for $R_0 < 6. 10^4$, the two other curves are obtained for horizontal turbulent diffusion coefficients equal to 3000 cm$^2$.s$^{-1}$ (dashed curve) and 15000 cm$^2$.s$^{-1}$ (dotted-dashed curve).}
\label{Dth}
\end{figure*}

\section{New derivation of the thermohaline mixing coefficient including radiative acceleration and/or external turbulence}

In this section we give a linear analysis of thermohaline convection in the presence of atomic diffusion and/or external turbulence. We begin with Boussinesq equations that describe motions in a nearly incompressible stratified viscous fluid, in the same way as D10, except that we add in the equations the terms due to atomic diffusion and turbulent diffusivity. 

Atomic diffusion is a selective process related to the fact that the stellar gas is composed of many different elements, each of them behaving its own way in the stellar gravity and radiation fields. If the radiative acceleration $g_R(i)$ acting on a given element (say iron) is locally larger than gravity, this element moves up with respect to the average stellar matter (mainly composed of protons). They move individually until they share their acquired momentum with the surroundings, which occurs after a collision time scale. The overall effect of this individual radiative acceleration on the medium is of order $g_R = \Sigma_i X_i g_R(i)$, where $X_i$ is the mass fraction of the considered element. The global radiative acceleration on the medium $g_R$ is generally several orders of magnitude smaller than the gravity, so that the medium as a whole is not destabilized (Figure 3). However, if the radiative acceleration on the considered element decreases with increasing radius, the diffusion flux decreases as well and this element accumulates in some layers inside the star. As a result, the internal stellar structure may be modified, due to changes in the opacity. Meanwhile, an inverse $\mu$-gradient is created, which may become unstable against thermohaline convection. 

External turbulence (for example rotation-induced mixing) may also coexist with thermohaline convection. In previous works (e.g. \cite {Charbonnel10}), such a situation was treated by simply adding the mixing coefficients of the two processes, computed separately, in spite of the analysis given by \cite{Denissenkov08}, who showed that the thermohaline mixing coefficient must be modified in this case, due to the action of the external turbulence on the falling blobs. Here we recalculate and confirm this effect and introduce it in the final expression of the thermohaline mixing coefficient.

Following D10, we first write the Boussinesq equations in a nearly incompressible fluid in the following form:
\begin{equation}
\frac{\partial \mathbf{v}}{\partial t}+(\mathbf{v},\nabla)\mathbf{v}=\frac{\delta \rho}{\rho_0}\mathbf{g_e}+\delta \mathbf{g_e}+\nu \nabla^2 \mathbf{v} 
\end{equation}
\begin{equation}
\frac{\partial T}{\partial t}+( \mathbf{v},\nabla) T=k^*_T \nabla^2 T
\end{equation}
\begin{equation}
\frac{\partial \mu}{\partial t}+(\mathbf{v},\nabla) \mu=k^*_{\mu} [\nabla^2 \mu + f(\mathbf{g_e}, T)]
\end{equation}

In the first equation, we have to take into account the fact that the gravity acting on the fluid is now an effective gravity defined as $g_e = g - g_R$, where $g_R$ is the overall radiative acceleration acting on the medium, namely $g_R = \Sigma_i X_ig_R(i)$. As already mentionned, $g_R$ is most generally negligible compared to $g$. However, when the effective gravities inside and outside the blobs are substracted, the $g_R$ terms are the only remaining ones, as they are different in the falling matter and in the surroundings.

The second equation, which describes the heat transfer between the falling blobs and the external medium, remains unchanged except that, in case of external horizontal turbulence, the heat transfer induced by the extra mixing must be taken into account in the thermal diffusivity. When the P\'eclet number, ratio of the horizontal turbulent diffusion coefficient to the thermal diffusivity is much smaller than one, $k^*_T$ reduces to  $k_T$, which is given by:
\begin{equation}
k_T=\frac{4 a c T^3}{3 \kappa c_P \rho^2}
\end{equation}
But when the P\'eclet number becomes larger than one, the horizontal diffusion coefficient must be added to $k_T$ in the expression of $k^*_T$.

The third equation concerns the diffusion of particles between the blobs and their surroundings. The coefficient $k_{\mu}$ is the so-called ``particle diffusivity" which, in hydrodynamical contexts, does not take into account the fact that stars are made of multicomponent gases. It is generally computed in an approximate way, and the values cited in hydrodynamical papers often correspond to the diffusion coefficient of helium in a hydrogen rich medium. Sometimes, it is even identified to the molecular viscosity, which is nothing else than the hydrogen-hydrogen self diffusion coefficient. In the atomic diffusion context, the diffusion coefficients for each element are computed in a consistent way (e.g. \cite{Paquette86}). It is not possible, in the present situation, to take into account the separate diffusion of all the elements through the blobs boundaries, but one must at least be careful to use for $k_{\mu}$ the diffusion coefficients corresponding to the most important elements (e.g. iron). The ratio of the diffusion coefficients for two different ions (1) and (2) of respective charges $Z_1$ and $Z_2$ is approximately $(Z_2/Z_1)^2$. Using that of iron leads to a smaller diffusivity than using that of helium, which would not be correct in this case (Figure 2). 

\begin{figure*}
\center
\includegraphics[width=0.5\textwidth]{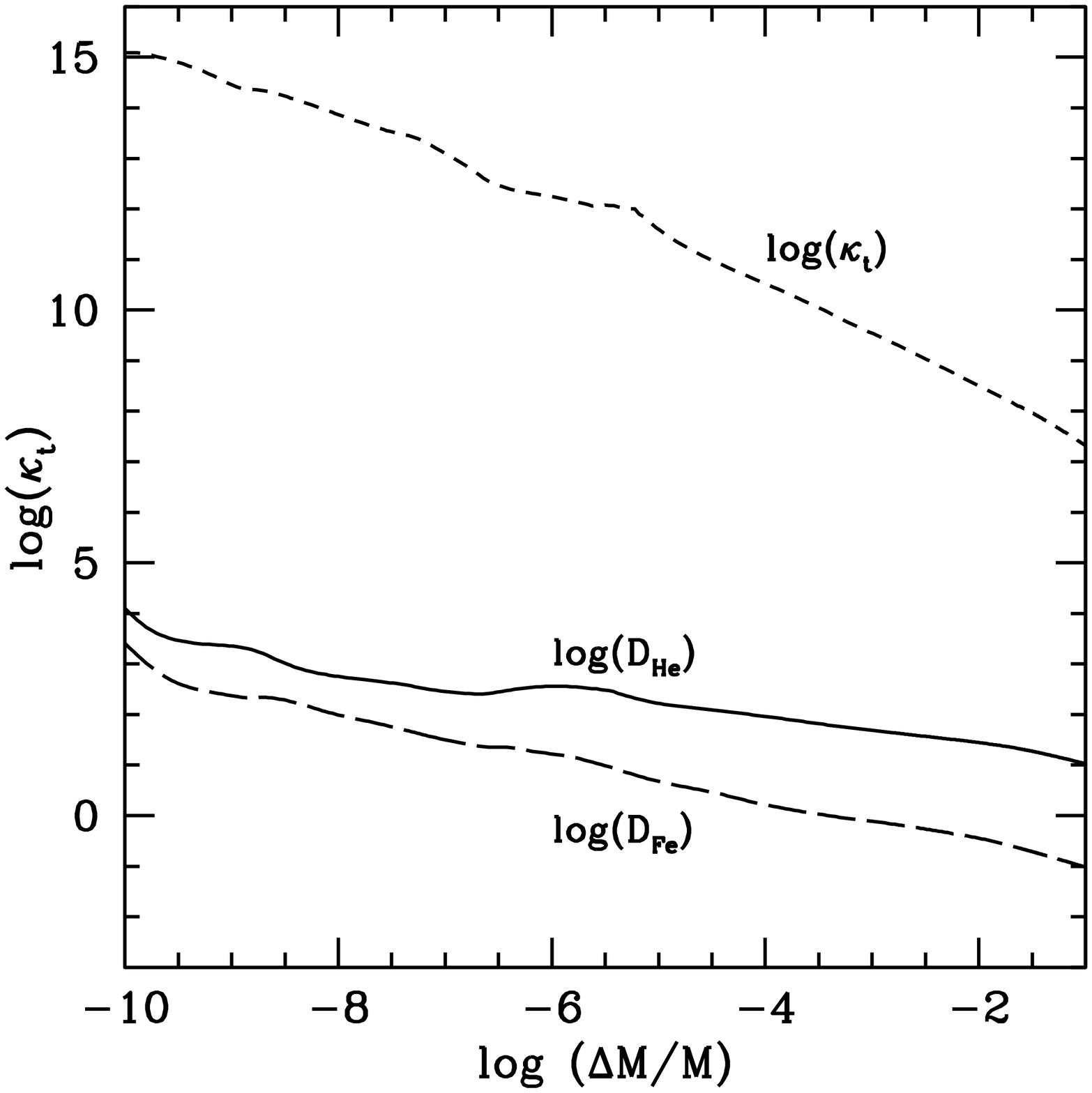}
\caption{Thermal diffusivity (short-dashed curve) and atomic diffusion coefficients (or particle diffusivities) for helium (solid curve) and iron (long-dashed curve) in the 1.7 M$_{\sun}$, 403 Myr model as a function of the external fractional mass.}
\label{grad}
\end{figure*}

In a general way, the particle diffusion between the interior and the exterior of the falling blobs is not only due to the abundance gradients, but also to the effects of gravity, radiative acceleration, temperature gradient, etc., which occur in the vertical direction (e.g. \cite{Vauclair82}). This is added as $f(\mathbf{g_e}, T)$ in Equation 7.  However, as pointed out by previous authors (e.g. D10), the vertical selective diffusion velocity is negligible compared to the downward velocity of the blobs so that we can ignore this effect. On the other hand, as mentionned above, the average effect of the radiative acceleration on the medium has to be taken into account in the computation of the velocity (Equation 5).

Finally, in case of external turbulence, the mixing of particles between the blobs and the inter-blobs is increased. The horizontal component of the extra turbulent diffusivity has then to be added to $k_{\mu}$ in equation 7. We have written the total diffusivity as $k^*_{\mu}$, which reduces to $k_{\mu}$ when no external turbulence is involved. 

In the following, we use the cartesian coordinate system with the vertical $z$ axis pointing in the direction opposite to the gravitational acceleration, the horizontal $x$ axis in a star's meridional plane and the horizontal $y$ axis perpendicular to them. Linearizing equations (5, 6, 7), with $w$ representing the vertical component of the blob velocity, we obtain the following system of equations, similar to equations 5 to 7 of D10 except for the gravity term and the effect of external turbulence included in $k^*_T$ and $k^*_{\mu}$:
\begin{equation}
\frac{\partial w}{\partial t}=g_e(\alpha \delta T -\beta \delta \mu)+ \nu \nabla^2 w + \delta g_e
\end{equation}
\begin{equation}
\frac{\partial \delta T}{\partial t}=-w \frac{\partial T}{\partial z}+k^*_T \nabla^2 \delta T
\end{equation}
\begin{equation}
\frac{\partial \delta \mu}{\partial t}=-w \frac{\partial \mu}{\partial z}+k^*_{\mu} \nabla^2 \delta \mu
\end{equation}
where $\alpha$ and $\beta$ are the usual coefficients of the equation of state:
\begin{equation}
\frac{\partial \rho}{\rho_0}=- \alpha\partial T + \beta\partial \mu
\end{equation}
Note that for simplicity we keep here the notation ${\partial T}/{\partial z}$ but we have to remember that in stars, due to the temperature stratification, this is equivalent to $(\nabla_{rad}-\nabla_{ad}) (T/H_P)$.

When a heavy blob falls down, the conditions inside it vary in a non adiabatic way, due to heat and particle diffusion with the surroundings. 
The vertical variations of $T$ and $\mu$ inside and outside the blob may be expressed in the following way:
\begin{equation}
\frac{\partial T}{\partial z}=\frac{\partial T_0}{\partial z} ( 1- \delta_T)
\end{equation}
\begin{equation}
\frac{\partial \mu}{\partial z}=\frac{\partial \mu_{0}}{\partial z} ( 1- \delta_{\mu})
\end{equation}
where the coefficients $\delta_T$ and $\delta_{\mu}$ have values between 0 (equality between the internal and external values of $T$ and $\mu$) and 1 (adiabaticity, no internal variations of $T$ and $\mu$). They may be defined from the following relations which give the differences $\delta T$ and $\delta \mu$ between the inside and the outside of the blobs.
 \begin{equation}
\delta T=- \frac{1}{2} \frac{\partial T_0}{\partial z} \ell \delta_T
\end{equation}
\begin{equation}
\delta \mu=- \frac{1}{2} \frac{\partial \mu_{0}}{\partial z} \ell \delta_{\mu}
\end{equation}
where $T_0$ and $\mu_0$ are the temperature and mean molecular weight outside the blobs and $\ell$ is the unknown length of the elongated cells (fingers).

The blobs stop falling when the buoyancy inside and outside are identical, that is when $\rho g_e$ are the same inside and outside.

As in D10, we transform equations 9, 10, 11 in terms of $w$, $\delta_T$ and $\delta_{\mu}$ and we search for solutions in the form of exp$(\sigma t)$ x exp$(i[k_x x + k_y y])$. But before that we have to express the new term $\delta g_e$ in the same way (in terms of $\delta_T$ and $\delta_{\mu}$ ). In the expression of the effective gravity, the only term which is different inside and outside the blobs is the radiative acceleration, which is a function of both the temperature and the chemical composition. Although these two variables are not completely disconnected, we will separate them in first approximation. This approximation will be discussed in the following section. We write:

\begin{equation}
\delta g_e= \frac{\partial g_e}{\partial T} \delta T + \frac{\partial g_e}{\partial \mu} \delta{\mu}
\end{equation}
or
\begin{equation}
\delta g_e=-\biggl(\frac{\partial g_e}{\partial T}\frac{\partial T_0}{\partial z} \frac{\ell}{2} \delta_T + \frac{\partial g_e}{\partial \mu} \frac{\partial \mu_0}{\partial z} \frac{\ell}{2} \delta_{\mu}\biggr) = -\biggl(\frac{\ell}{2}  \frac{\partial \mu_{0}}{\partial z}  [\frac{\partial g_e}{\partial T}  R_0 \frac{T}{\mu}\delta_T +  \frac{\partial g_e}{\partial \mu} \delta_{\mu}]\biggr)
\end{equation}

In the following, for simplicity we assume a perfect gas law, so that $\alpha = 1/T$ and $\beta = 1/\mu$. Also, for simplicity we write here $k_T$ and $k_\mu$ instead of $k^*_T$ and $k^*_\mu$. Defining $g_{e,T}$ as $\partial \ln g_e/\partial \ln T$ and $g_{e,\mu}$ as $\partial \ln g_e/\partial \ln \mu$,
the preceding equations may then be reduced to :
\begin{eqnarray}
\lefteqn{4 \sigma^3 +2(k_T+k_{\mu}+2 \nu)k^2 \sigma^2 + } \nonumber\\
& & \left\{[k_T k_{\mu}+2 \nu(k_T+k_{\mu})]k^4 
+ 2 g_e \frac{\partial ln\mu_0}{\partial z}(R_0-1)\right\} \sigma \nonumber \\
& & - g_e \frac{\partial ln\mu_0}{\partial z}(k_T-R_0 k_{\mu}) k^2 +\nu k_T k_{\mu} k^6 \nonumber\\
& &+2\sigma g_e \frac{\partial ln\mu_0}{\partial z}(g_{e,\mu} +  R_0 g_{e,T}) + k^2 g_e \frac{\partial ln\mu_0}{\partial z} (k_T g_{e,\mu} + R_0 k_{\mu} g_{e,T}) =0
\end{eqnarray}
with $k^2 = k^2_x + k^2_y$

Taking into account that the thermal diffusivity is much larger than the particle diffusivity and the molecular viscosity, we neglect the same terms as D10 and this equation is simplified in:

\begin{equation}
\sigma= \frac{k^2}{2} \frac{(k_T - R_0 k_{\mu}) + (k_T g_{e,\mu} + R_0 k_{\mu}g_{e,T})}{(R_0 - 1) - (g_{e,\mu} + R_0 g_{e,T})}
\end{equation}

The thermohaline mixing coefficient, of the order of $1/2 \sigma\ell^2$, becomes:
\begin{equation}
D_{th}=C_t k^*_T \frac{(1 - R_0 \tau^*) + (g_{e,\mu} + R_0 \tau^* g_{e,T})}{(R_0 - 1) - (g_{e,\mu} + R_0 g_{e,T})} 
\end{equation}
where $k_T$ has been replaced by $k^*_T$ and the inverse Lewis number $\tau$ has been replaced by $\tau^* = k^*_{\mu}/k^*_T$. $C_t$ is the usual constant related to the aspect ratio of the fingers, that we adjust using TGS11 simulations.

If the derivatives of the effective gravity are neglected, as well as the horizontal turbulence, this expression of the thermohaline mixing coefficient is identical to the D10's one. If only horizontal turbulence is introduced, this expression is similar to that obtained by \cite{Denissenkov08}.

In the general case, the two limits for the development of thermohaline convection are slightly modified. The boundary with dynamical convection, which occurs for large $\mu$-gradients, is not exactly $R_0 = 1$ but $R_0 = (1 + g_{e,\mu})/(1-g_{e,T})$. The other boundary, for small $\mu$-gradients, becomes $R_0 = (1+g_{e,\mu})/[\tau^*(1-g_{e,T})]$ instead of $R_0 = 1/\tau$ . The orders of magnitude of these corrections are discussed in the next section.

\section{Orders of magnitude and discussion}

In the previous section, we have introduced the partial derivatives of the effective gravity $g_{e,\mu} = \partial \ln g_e/\partial \ln \mu$ and $g_{e,T} = \partial \ln g_e/\partial \ln T$. We now have to understand their real meaning, and derive their orders of magnitude in stellar situations. In the following we first assume no external turbulence. We will discuss the effect of adding turbulence later on.

The effective gravity is $g_e = g - g_R$ where $g_R$ is the radiative acceleration transferred to the whole medium by the elements which are individually levitated, namely
$g_R = \Sigma_i X_i g_R(i)$ where $X_i$ is the mass fraction of the element $i$ and $g_R(i)$ the radiative acceleration acting on the same element $i$. In the following, to evaluate the orders of magnitude of these terms, we will simplify the problem by assuming that one major element is the reason for the $\mu$-gradient inversion (e.g. iron). Generalisation to several elements will be done numerically in stellar evolution codes. 

Even in the case of one levitating element, the situation is not simple as each ion behaves in a different way, which is precisely the reason for the selective accumulation.
In this case we write $g_R = X_i g_R(i)$ where $g_R(i)$ represents the contribution to the radiative acceleration of all the ions of the considered element, that is:
\begin{equation}
g_R(i) = \Sigma_j X_{i,j} g_{i,j} = X_i \Sigma_j \frac{X_{i,j}}{X_i} g_{i,j} 
\end{equation}
where $j$ represents the various ionisation stages of the element $i$.

The variation of $g_R$ with $\mu$ is a combination of two effects. First the relative mass fraction of the heavy element $X_i$ appears as a multiplicating coefficient in $g_R$. Second this mass fraction has also an influence on $g_R(i)$ through the saturation of the lines. Namely, when the abundance of the element increases, the effect of the saturated lines on the total radiative acceleration of the considered element decreases because all the considered ions have to share the same available photon flux. 

The variation of $\mu$ induced by the accumulation of element $i$ may be written:
\begin{equation}
\delta \ln \mu \approx \delta{X_i} [1 - \frac{\mu \bar{n_i}}{A_i}] 
\end{equation}
where $\bar{n_i}$ is the average number of particles (including electrons) associated to element $i$ in the prevailing physical conditions and $A_i$ its mass number.

For the considered heavy elements, the second term is small compared to one (otherwise there would not be any inverse $\mu$-gradient). We neglect it in first approximation and write $g_{e,\mu}$ as:
\begin{equation}
g_{e,\mu} = \frac{\partial \ln g_e}{\partial \ln \mu} \approx \frac{1}{g} \frac{\partial g_R}{\partial X_i} = \frac{g_R(i)}{g} + \frac{X_i}{g} \frac{\partial g_R(i)}{\partial X_i}
\end{equation}
In this equation, the first term is positive. The second term is negative, difficult to evaluate, but its maximum importance is reached if $g_R(i)$ varies like $1/X_i$ , in which case  $g_{e,\mu}$ vanishes. On the contrary, if the saturation effects are neglected, the $g_{e,\mu}$ term can reach the order of magnitude of $g_R(i)/g$, which may become larger than one (Figure 3).

\begin{figure*}
\center
\includegraphics[width=0.5\textwidth]{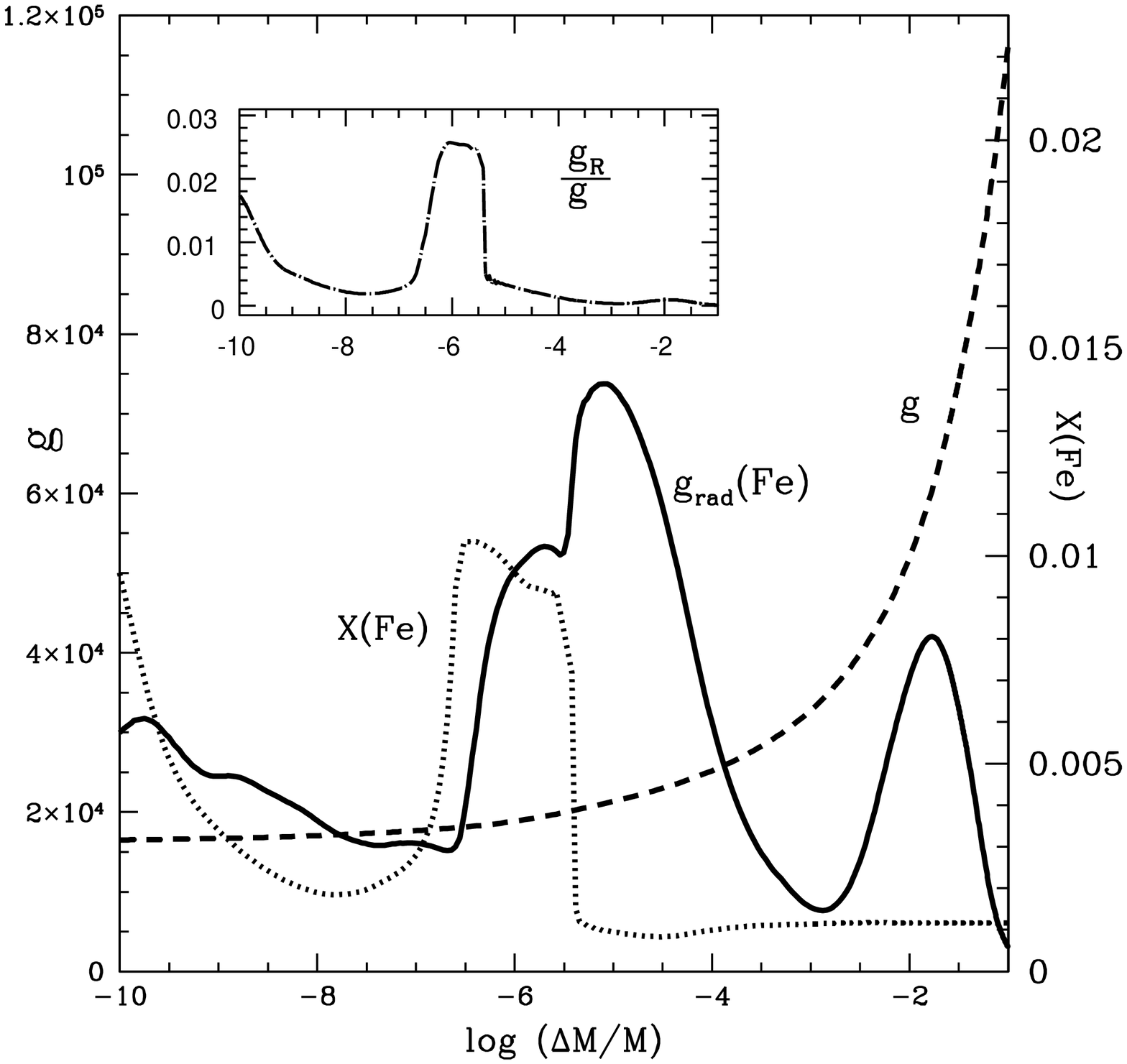}
\caption{Radiative acceleration on iron (large frame) and total radiative acceleration on the stellar medium (small frame) compared with the local gravity in a $1.7 M_{\sun}$, 403 Myr model. The dashed curve in the large frame represents the gravity, plotted as a function of the external fractional mass. The solid curve is the radiative acceleration on iron, which largely exceeds the gravity in several stellar layers, leading to the iron accumulation. The dotted curve represents the mass fraction of iron in this model, showing the radiatively-induced accumulation. The small frame presents, with the same absissa, the ratio of the global radiative acceleration ($g_R = X(Fe) g_{rad}(Fe)$) induced by iron on the stellar medium, to the local gravity. It never exceeds $3\%$.}
\label{gradb}
\end{figure*}

The variation of $g_R$ with temperature only appears through $g_R(i)$ and comes mostly from the fact that the various ions of the same elements are not pushed up in the same way by the radiative acceleration due to their different atomic structure:
\begin{equation}
g_{e,T} = \frac{\partial ln g_e}{\partial ln T} = \frac{T}{g} \frac{\partial g_R}{\partial T} = \frac{T}{g} X_i \frac{\partial g_R(i)}{\partial T}
\end{equation}
or:
\begin{equation}
g_{e,T} = \frac{T}{g} X_i \Sigma_j \biggl(X_{i,j} \frac{\partial g_{i,j}}{\partial T} + g_{i,j}\frac{\partial X_{i,j}}{\partial T}\biggr)
\end{equation}

The order of magnitude of the first term may be evaluated with the assumption that $g_{i,j}$ varies inside the star approximately like $1/T$ \citep{Michaud76}. It is negative, of order $-g_R/g$.
The second term is more difficult to evaluate, as $X_{i,j}$ depends on the Saha equation. It may either increase or decrease with T. In the approximation of two ionisation stages, $j$ and $(j-1)$, the fraction of ion $j$ is approximately proportional to $exp(-E_{j-1}/k_BT)$, where $E_{j-1}$ is the ionisation energy of ion $(j-1)$, so that 
its contribution to the partial derivative of $g_e$ may be approximately evaluated as $(g_R/g) (E_{j-1}/k_B T)$, which may become larger than $g_R/g$. In our reference model (1.7 M$_{\sun}$ at 403 Myr), this term never exceeds $15\%$ whereas $g_R/g$ always remains below $3\%$. 

It is interesting to note that, due to the signs of these terms, the effect of the radiative acceleration on the thermohaline mixing coefficient is often to reduce it, but may also sometimes increase it. In the regions where the radiative acceleration decreases with increasing radius, the falling blobs suffer less levitation than the surroundings, thereby falling more easily. In any case, these terms need to be precisely tested in the computations of stellar models.

Let us now discuss the effect of an external turbulence on the thermohaline convection. As already pointed out by \cite{Denissenkov08}, the presence of horizontal turbulence decreases the efficiency of the thermohaline mixing and even suppresses it completely as soon as the horizontal turbulent diffusivity is larger than the thermal diffusivity, in other words as soon as the horizontal P\'eclet number is larger than one. In the small frame of Figure 1, the curves represent the thermohaline coefficients computed by including horizontal turbulent mixing coefficients of 3000 cm$^2$.s$^-1$ (dotted curve) and 15000 cm$^2$.s$^-1$ (dotted-dashed curve). These turbulent diffusion coefficients are small, but the consequences on the thermohaline convection is important.The basic effect is that the thermohaline coefficient drops for values of $R_0$ decreasing with increasing turbulence. This means that thermohaline convection needs a larger $\mu$-gradient to take place.
If the external turbulence is due to rotational mixing as modelled by \cite{Zahn92} and coworkers, the importance of the horizontal mixing, much larger than the vertical one, cannot be neglected. As discussed by \cite{Zahn92}, whereas the P\'eclet number is always smaller than one in the vertical direction, it may become high and even larger than one in the horizontal direction. In this case, simply adding the turbulent diffusion coefficients describing the thermohaline convection and the rotation-induced mixing strongly overestimates the global mixing efficiency.

\section{Conclusion}

We have derived a new expression for the thermohaline diffusion coefficient in stars, which occurs in case of mean molecular weight inversion, taking into account the influence of radiative accelerations and external turbulence. We find that the corrections induced by radiative accelerations are small but should not be neglected in the computations of stellar models. When relatively abundant elements like iron are strongly pushed up by radiation, the momentum transfer to the surroundings leads to a negligible effect on the stellar gas as a whole. The only important influence of radiative accelerations on thermohaline convection is due to the difference in the radiatively induced momentum inside the fingers (falling blobs) and outside (surrounding medium). This difference is small compared to the other effects, but in any case the influence of radiative accelerations on the thermohaline efficiency should be systematically checked in all computations of thermohaline convection induced by the radiative levitation of specific elements.

These results are important as they confirm the idea that the accumulation of radiatively-levitated elements in stars leads to mixing in a straightforward way. Up to now, atomic diffusion including selective radiative accelerations was known to qualitatively account for a large number of stellar observations (e.g. chemically peculiar stars). However, atomic diffusion alone leads to abundance variations much too important for a quantitative fit to the abundance determinations. Macroscopic motions like rotation-induced mixing, mass loss, etc. had to be invoked to try explaining the observations. Here we show that a fundamental process was forgotten in all previous computations: thermohaline instabilities directly induced by the atomic diffusion process itself. All the computations of radiative levitation of chemical elements will have to be revised, taking this effect into account. Note that in the case where heavy elements are levitated, helium always sinks due to gravitational settling, as the radiative flux is never able to sustain it in its original abundance. This creates a stabilizing $\mu$-gradient which opposes the destabilizing one due to the heavy elements. All these effects have to be treated together, and the expected result is that heavy elements may still accumulate in connexion with the helium depletion, but not as much as obtained with pure diffusion (C.f. \cite{Theado09}).

Furthermore, these processes have to compete with external turbulence and particularly rotation-induced mixing. This will be somewhat complex, as it depends on the ratio of the horizontal to the vertical mixing coefficients. We have shown that a large horizontal turbulence decreases and may even suppress the efficiency of the thermohaline mixing. On the other hand, vertical turbulence increases this efficiency. A combination of all these effects, including their close interaction, will have to be studied in real stellar cases.

\acknowledgments
SV acknowledges fruitful discussions with Dr. Haili Hu at the Kavli Institute for Theoretical Physics, University of California, Santa Barbara, during the program "Asteroseismology in the Space Age", September to December 2011.

\end{document}